# Proposal that interpretation of field emission current-voltage characteristics should be treated as a specialized form of electrical engineering

Richard G. Forbes[a)]

University of Surrey, Advanced Technology Institute & School of Computer Science and Electronic Engineering, Guildford, Surrey, GU2 7XH, UK

[a)]Electronic mail: r.forbes@trinity.cantab.net

This article proposes we should think differently about predicting and interpreting measured field electron emission (FE) current-voltage [$I_m(V_m)$] characteristics. It is commonly assumed that $I_m(V_m)$ data interpretation is a problem in emission physics and related electrostatics. Many experimentalists then analyze data using Maxwell's electrostatics and a methodology (the Fowler-Nordheim plot) developed in the late 1920s. However, with modern emitting materials this 90-year-old interpretation methodology often fails (maybe in nearly 50% of cases): it yields spurious values for characterization parameters, particularly field enhancement factors (FEFs). This has generated an unreliable literature where industrialists or defence scientists interested in FE applications need to treat ALL published FEF values as unverified, unless proved otherwise, for example by one or more validity checks. A new validity check, that supplements existing checks, is described. Twelve different "system complications" that, acting singly or in combinations, can cause validity-check failure are identified. A top-level path forwards from this unsatisfactory situation is proposed, as follows. The term *field electron emission system*

*(FE system)* is defined to include all aspects of an experimental system that affect the measured current-voltage characteristics. This includes: emitter composition, geometry and surface condition; geometrical, mechanical and electrical arrangements in the vacuum system; all aspects of the electronic circuitry and all electronic measurement instruments; the emission physics; the system electrostatics; and all other relevant physical processes. The science and engineering of FE systems is to be regarded as a specialized form of electronic/electrical engineering, provisionally called *FE Systems Engineering*. In this approach, the $I_m(V_m)$ relationship is split in two: (a) the current is expressed as a function $I_m(F_C)$ of the local surface-field magnitude $F_C$ at some defined location "C" near the emitter apex (or, in multi-emitter systems, the apex of the sharpest emitter); and (b) the relationship between $F_C$ and the measured voltage $V_m$ is written in one of the equivalent alternative forms $F_C = V_m/\zeta_C$ or $F_C = K_C V_m$, where $\zeta_C$ and $K_C$ are defined by these equations. Determining $I_m(F_C)$ is a problem in emission physics; determining the behavior of $\zeta_C$ (or $K_C$) involves the system electrostatics and (for systems failing a validity check) other aspects of FE Systems Engineering. Inadequate research attention has been given to the behavior of $\zeta_C$ (or $K_C$) for systems failing validity checks due to "system complications". A problem is that several different "system complications" may operate simultaneously. An unsolved research problem is to diagnose which one or more of these is/are operating in any given case. Tentative outline suggestions are made about how this might be done. Much more research is needed. It might also be useful to develop a short course in FE systms engineering and $I_m(V_m)$ data interpretation and more generally in FE systems engineering: suggestions are made about possible content. Finally, it is argued that thinking in the way described would be a useful step towards putting the topic of field electron emission onto a better and more respectable scientific basis, and that the International Vacuum Nanoelectronics Conference "Call for Topics" could usefully be modified.

## I. INTRODUCTION

### A. *General introduction*

Large-area electron emission sources have many actual and potential technological applications. One available source-technology is field electron emission (FE). In consequence, the last twenty-plus years have seen extensive amounts of related emitter-materials research. This has mostly aimed at developing large-area field electron emitters (LAFEs) that have high macroscopic (i.e., "area-average") electron-current density $J_M$. (This article uses the customary field emission convention that electron currents and current densities are treated as positive, although negative in classical electromagnetism.)

Very many papers relating to FE technology make the assumption that the interpretation of *measured* current-voltage [$I_m(V_m)$] characteristics is a problem in emission physics and related (zero-current) electrostatics. Such papers often then analyze their data using the electrostatics of 150 years ago[1] and a data-analysis methodology—the Fowler-Nordheim (FN) plot—introduced[2] in 1929, nearly 100 years ago.

The main objective of most FN-plot users has been the extraction of emitter electrostatic characterization data from the FN-plot slope. For this objective, data analysis based on FN plots worked adequately in the period 1929 to around 1950 (or slightly later), because in this period (and earlier in the 1920s): (a) field emitters were nearly always metal; (b) mounting arrangements nearly always provided a low-electrical-resistance path to the high voltage source; (c) emitters were usually operated at fields where space-charge effects were not a major influence; and (d) system vacuum levels were good enough that vacuum breakdown was not a significant factor. Thus, in this period, FE systems were nearly always what is described below as *electronically ideal*.

With modern emitting materials (semiconductors, carbon-based materials, etc.), which began to be seriously developed from around the 1950s, the FE-system characteristics just stated are no longer necessarily all applicable. As a result, in many modern cases (maybe in nearly 50% of cases) the 90-year-old data-analysis methodology based on FN plots fails: it can yield spurious values for emitter characterization parameters, particularly field enhancement factors (FEFs)[3]. Improved modern FE data-analysis techniques, such as the Murphy-Good plot[4] and numerical multi-variable regression techniques[5] based on the Kyritsakis-Xanthakis ("Earthed sphere") FE equation[6], can also fail if they are applied to emitters and FE systems that are not electronically ideal.

In the author's perception, this situation has led to massive breakdown of the peer-review system in this technological area, and has generated a FE-technology literature where sensible users of published FE characterization data (but particularly industrialists interested in medical applications and national defense scientists) should treat ALL published FEF-values as UNVERIFIED, unless or until results of interest have been or can be proved valid—e.g., by some form of *validity check* (see below).

This paper proposes that we should *think differently* about how to predict and interpret measured FE current-voltage characteristics. In principle, we should treat the analysis of measured FE current-voltage characteristics as a specialized form of electrical and electronic engineering. Perhaps this could be called *FE Systems Engineering.*

FE Systems Engineering would not be ordinary electrical/electronic engineering, because it would involve currents, voltages *and fields* (and more besides), but this is also true of some other forms of electrical engineering, for example high-voltage engineering. A field electron emitter is, in effect, a new form of electronic circuit element, and appropriate field-emitter-specific forms of electrical circuit theory will probably need to be developed.

This paper aims to present a top-level overview of the main issues that seem to arise when attempting to establish FE Systems Engineering. This overview includes brief discussion of some specific issues that have already been discussed in more detail elsewhere. The paper itself is based on a presentation made at the 2022 35th International Vacuum Nanoelectronics Conference[7] (see: https://doi.org/10.13140/RG.2.2.10294.57927) , but aspects of the argument have been developed further since the original presentation.

The structure of the rest of the paper is as follows. After a brief description In Section IB of how FN plots are currently used, Section II identifies twelve "system complications" that can cause FN-plot-based data-analysis methodology to fail. Section III then outlines the top-level components of a new, more general approach. Section IV discusses existing validity checks and a new form of check. Section V discusses the need to develop methods for diagnosing the cause or causes of validity-check failure, and makes some initial suggestions about how this might be done. Section VI makes some other suggestions related to the development of FE Systems Engineering. Section VII provides a summary. An Appendix makes suggestions for possible content of a short course related to FE Systems Engineering.

## B.  1929-style FE data-analysis methodology as currently used

To provide a starting point, this Section describes 1929-style FE data-analysis methodology[2] as currently commonly used. As is well known, the commonest (but not the only possible) experimental FE data-analysis methodology is the Fowler-Nordheim (FN) plot, which—when natural logarithms are used—is a data plot of the form $\ln\{Y/X^2\}$ vs $1/X$, where $X$ is the input variable (normally a field or a voltage) and $Y$ is the output variable (normally a current or a current density). The "curly bracket notation" $\{Q\}$ means "take the numerical value of $Q$ when $Q$ is expressed in the designated units".

Nowadays, FN and related data-analysis plots are most commonly made using natural logarithms. Although common logarithms have been used in older work, and are sometimes still used, it is usually better practice to convert all common logarithms to natural logarithms. Thus, only plots using natural logarithms will be discussed here.

Obviously, when raw measured current-voltage $I_m(V_m)$ data are used, then $X$ becomes $V_m$, and $Y$ becomes $I_m$. The author's view is that using raw measured $I_m(V_m)$ data is best practice, because the resulting FN or related plot is then clearly an experimental result. The local field magnitude $F_C$ at some characteristic location "C" on the emitter surface, at or near its apex can then be related to $V_m$ by the equation

$$F_C = V_m/\zeta_{mC}, \tag{1}$$

where $\zeta_{mC}$ is a characteristic *local voltage conversion length (LVCL)*, defined by this equation.

An *electronically ideal* FE system is, by definition, one where the parameter $\zeta_{mC}$ is constant, independent of the values of $V_m$ and $I_m$, and the measured current-voltage characteristics are determined by the emission physics and system electrostatics alone. For an electronically ideal FE system, an experimental FN plot of type $\ln\{I_m/V_m^2\}$ vs $1/V_m$ is "nearly straight" and $\zeta_{mC}$ is a constant that can be extracted from the slope of the plot. (Or, alternatively, the reciprocal of $\zeta_{mC}$ can be used in the theory.)

However, particularly when analyzing data from LAFEs, where the individual emitters are "protrusions" that stand on an underlying substrate, it has become customary to *pre-convert* the experimental data, in one or both of the following ways, before making a data-analysis plot.

First, a *macroscopic (or LAFE-average) current density $J_M$* can be defined by

$$J_M = I_m/A_M, \tag{2}$$

where $A_M$ is the *macroscopic area (or "footprint")* of the emitter. The explicit inclusion of the subscript "M" on $J_M$ is important, because massive confusion exists in FE technological literature between $J_M$ and characteristic local emission current density $J_C$ (both of which are commonly denoted by the same symbol $J$). $J_C$ is much larger than $J_M$.

Second a so-called *macroscopic field $F_M$* (also called an "applied field") is defined within the context of the relevant system geometry. This field is the electrostatic field magnitude at the position (on the substrate) of the base of the protrusion, in the absence of the protrusion. It can be written in terms of $V_m$ by the equation

$$F_M = V_m/d, \tag{3}$$

where $d$ is a parameter with the units of distance. Many LAFEs use so-called "PPP" geometry where the protruding emitters are on one of a pair of parallel planar plates of lateral extent that is large in comparison with the plate separation. If a system with this geometry is electronically ideal then $d$ is a constant equal to the plate separation. If the system is not electronically ideal then $d$ will likely be a variable parameter not equal to the plate separation.

With LAFEs, it is customary to define a characteristic macroscopic field enhancement factor (MFEF), denoted here by $\gamma_{MC}$ (but often by $\beta$ in FE literature) via the relation

$$\gamma_{MC} \equiv F_C/F_M = d/\zeta_{mC}, \qquad (4)$$

where the second part of the relation follows from eqns (1) and (3). For electronically ideal emitters, $\gamma_{MC}$ is a constant: when the protrusion height is very much less than $d$, $\gamma_{MC}$ is a useful parameter that characterizes the sharpness of the most strongly emitting protrusions.

When the measured voltage has been pre-converted to a macroscopic field, then (for electronically ideal emitters) the value of $\gamma_{MC}$ can be obtained directly from the slope of a FN or related plot made by taking $X$ as $F_M$, and $Y$ as either $J_M$ or $I_m$. Alternatively (and better in the author's view[8]), again for electronically ideal emitters, $\zeta_{mC}$ can be obtained from the slope of FN or related plot made using the raw measured $I_m(V_m)$ data, and $\gamma_{MC}$ then obtained as $d/\zeta_{mC}$.

In summary, an electronically ideal FE system is one for which the parameters $\zeta_{mC}$, $d$, and $\gamma_{MC}$ are constant independent of the measured voltage and measured current. For such systems, values of these parameters can be obtained from the slope of a FN or related data-analysis plot. For such systems, values of area-like quantities can be obtained from the intercept that the experimental plot makes on the vertical (1/$X$=0) axis (see

Ref. 4), but procedures are not discussed in detail here. The underlying point is that 1929-style FE data analysis works reliably only for electronically ideal systems; it fails in many modern cases, where the FE system is not ideal.

## II. CAUSES FOR BREAKDOWN OF 1929-STYLE DATA-ANALYSIS METHODOLOGY

Even for electronically ideal emitters and FE systems, the 1929 FN plot is not the most modern or the most useful form of FE experimental data-analysis now available. However, a much bigger data-analysis issue is that many reasons are now known that cause FE systems *not* to be electronically ideal, and more reasons continue to be realized. These causes have been termed *system complications*. As of the time of submission (January 2023), the following are the known possible system complications.

(*1) Significant series resistance in the current path between the emitter and the high-voltage generator. This cause includes safety resistors and current-measurement resistors whose circuit effects have not been correctly taken into account, and series transistors used as current-limiting devices.

(*2) Voltage-loss effects along the emitter (which can lead to current-dependence in characterization parameters such as $\zeta_{mC}$ and $\gamma_{MC}$).

(*3) Effects due to field emitted vacuum space charge (FEVSC).

(*4) Field-dependent changes in FE system geometry (reversible or permanent) due to Maxwell stress, and permanent changes due to various forms of emitter erosion.

(*5) Changes in local work function due to current-related heating effects (Joule and/or Nottingham heating) and related desorption of surface adsorbates.

(*6) Various effects related to field penetration into semiconductor emitters, and/or to the condition of the semiconductor surface, and/or to the mechanism of electron supply.

(*7)   Charge-trapping on non-metallic surfaces adjacent to the emitting surface.

(*8)   Use, in data analysis, of a seriously incorrect value of the effective emitter local work function.

(*9)   The actual operating regime of the emitter is not the Murphy-Good FE regime.

(*10)  The emitter is so sharp that Murphy-Good FE theory does not apply.

(*11)  With so-called Nanoscale Vacuum Channel Transistors (NVCTs), use of diode-type theory that does not accurately model the behavior of a three-terminal electronic device.

(*12)  With LAFEs, plot non-linearity (upwards bending at the left-hand-side of a FN plot) can occur because the emission comes from a distribution of many individual emitters, all with different field enhancement factors.

The situation is made even more complicated if (as is sometimes, or maybe often, the case) several different system complications are operating simultaneously.

## III. TOP-LEVEL COMPONENTS OF A NEW APPROACH

The following sub-sections outline top-level components of a proposed new, more-general, approach.

### A.   *Field electron emission systems*

The term *field electron emission system (FE system)* is defined to include all aspects of an experimental or technological system that affect the measured current-voltage characteristics. This includes:

- the emitter *configuration* (composition, geometry and surface condition);
- geometrical, mechanical and electrical arrangements in the vacuum system;
- all aspects of the electronic circuitry and all electronic measurement instruments;

- the emission physics;
- all other relevant physical processes that might be happening (as exemplified by some items in the list in Section II).

As already indicated, the proposal is that the science and engineering of FE systems should be regarded as a specialized form of electronic/electrical engineering, called here FE Systems Engineering.

## B.  The components of FE Systems Engineering

The theory of FE Systems Engineering has two main components: Emission Physics; and Circuit and Field-Voltage (CFV) Behavior. Field Emitter Electrostatics can be seen as an auxiliary topic that contributes to both, but more to CFV behavior.

*The Emission Physics* yields an expression for the predicted emission current $I_e$ as a function of the local electrostatic field magnitude $F_C$ at some characteristic location "C" near or at the emitter apex "a". Thus, the emission physics yields $I_e(F_C)$. If there is no significant leakage current, then this can be treated as a *predicted measured current* $I_m(F_C)$. (If there is significant leakage current, then a correction factor needs to be included[9].)

Modern accounts of basic FE emission physics can be found, for example, in Refs. 10–12.

Note that, due to uncertainties in (a) current understandings of emission physics and (b) the accuracy of models for the emitter configuration, predictions of measured current for real emitters are subject to massive uncertainty[9], perhaps sometimes as much as a factor of 1000.

At least in principle, analyzing the *Circuit and Field-Voltage (CFV)* aspects of FE system behavior provides a link between the characteristic local-field magnitude $F_C$ and the measured voltage $V_m$. As already indicated, this link can be written in form (1) above. Alternatively, it can be written in the equivalent form

$$F_C = K_{mC} V_m, \tag{5}$$

where $K_{mC}$ [$\equiv 1/\zeta_{mC}$] is the *(characteristic) voltage-to-local-field conversion factor*. In both cases the "voltage" in question is the measured voltage, but (for simplicity) this is not explicitly indicated in the terminology, in this paper. Typically, $\zeta_{mC}$ is measured in "nm", but $K_{mC}$ in "m$^{-1}$".

[Note that the conversion factor $K_{mC}$ is usually denoted by $\beta$ (or $\beta_V$) in FE literature, but I avoid this notation, in order to prevent confusion with the dimensionless field enhancement factor commonly denoted by $\beta$ in FE literature.]

In reality, in nearly all cases, each of the individual "FE system complications" has its own "theory of $\zeta_{mC}$ (or $K_{mC}$)". No integrated general theory of CFV behavior (i.e., no general theory of $\zeta_{mC}$ or $K_{mC}$) currently exists.

In a few cases, integrated theory of two system complications has been developed — e.g. "series resistance" and "voltage loss" have been jointly discussed[13]. However, in many cases it can be that, due to lack of appropriate theory, experimental current-voltage characteristics cannot be reliably interpreted.

In practice, this implies an engineering need to first sort measured current-voltage [$I_m(V_m)$] into two groups into: those that can be reliably interpreted (if up-to-date emission theory is used); and those that—at present—cannot. This initial sorting process can be formalized as described next.

## C.   *Electronically ideal FE systems and Orthodoxy*

As indicated earlier, an FE system is termed *electronically ideal* (in a range of $V_m$–values of interest) if no "system complications" are operating in the voltage-range of interest, and hence the the parameters $\zeta_{mC}$ and $K_{mC}$ are constant, independent of the values of $V_m$ and $I_m$. With LAFEs, the parameters $d$ and $\gamma_{MC}$ will also be constant. The behavior of an electronically ideal system can be described by emission physics and zero-current system electrostatics alone.

An electronically ideal FE system is further described as *orthodox* (or as orthodoxly behaving) if the emission process can be adequately described by Murphy-Good FE theory[14] (for a modern derivation see Ref. 15), and if the value of the relevant emitter local work-function (assumed unchanging during the relevant experimental run) is adequately known. An FE system may be "orthodoxly behaving" over part of its $V_m$ operating range (usually the low-voltage part) but may be "non-orthodox" in other parts of the range.

For FE systems behaving orthodoxly, traditional data-analysis methodology works "more-or-less adequately" (if up-to-date emission theory is used). However, as discussed elsewhere[16], improved modern versions of data-analysis theory (e.g., the Murphy-Good plot[4,16]) may often work slightly better.

## IV. VALIDITY CHECKS AND THE ORTHODOXY TEST

### A. Overview

The term *validity check* refers to a check that FE system behavior is orthodox and hence that characterization parameters extracted using traditional methodology are adequately valid (if up-to-date emission theory is used). Three forms of validity check are now recognized.

(1) *Near-linearity of a Fowler-Nordheim (FN) plot (or similar data-analysis plot).* The theory and analysis of FN and related data plots has recently been discussed elsewhere[4,16].

(2) *The Orthodoxy Test.* This is well described in the literature[3,16]. Essentially, the procedure tests, for the given emitter, whether the apparent operating range of characteristic local fields $F_C$, as derived from a FN or similar data plot is physically reasonable, as compared with well-established operating ranges for metal emitters. The test is in fact carried out with the quantity *characteristic scaled field $f_C$*, which is related to $F_C$ by $f_C = F_C / F_R$, where the *reference field $F_R$* is the field necessary to pull down to the Fermi level the top of a Schottky-Nordheim barrier of zero-field height equal to the local work function $\phi$.

*(3) The "Magic Emitter" test.* This is a new test, described in the next subsection.

As indicated above, in the FE materials-technology community, it is common practice, before making a data-analysis plot, to "pre-convert" experimental $I_m(V_m)$ data into an alternative form involving macroscopic fields. This is done by using a conversion equation that—in fact[8]—is often of questionable validity, and hence may yield "apparent macroscopic fields" rather than "true macroscopic fields". Tests (1) and (2) above will normally work with "pre-converted" data-sets, but it is strongly advised that it is better practice to use the raw measured $I_m(V_m)$ data to make the data-analysis plot and carry out validity checks of the first two types.

## B. The "magic emitter" test

The magic emitter test is a validity check that can be carried out on published experimentally derived data, when this is plotted in a pre-converted form that shows the range of apparent macroscopic fields $F_M^{app}$ over which significant FE current is emitted, and when the relevant local work function value $\phi$ is stated (or can be estimated from the stated nature of the emitting material).

Papers of interest will almost always report the value of the apparent field enhancement factor ($\gamma_{MC}^{app}$ here, but usually "$\beta$" in FE literature). From eq. (3) above, using this value $\gamma_{MC}^{app}$, the stated range of apparent macroscopic fields $F_M^{app}$ can be converted to a corresponding range of apparent characteristic local fields $F_C^{app}$.

For a material of local work-function $\phi$, the corresponding reference field $F_R$ as defined above is given by (see: https://doi.org/10.13140/RG.2.2.34321.35681/1 )

$$F_R = (4\pi\varepsilon_0/e^3) \phi^2 \cong (0.6944615 \text{ V/nm}) \cdot (\phi/\text{eV})^2 . \quad (6)$$

This value can be used to convert the range of apparent characteristic local fields $F_C^{app}$ to a corresponding range of values of apparent characteristic scaled-field $f_C^{app}$, using the equation

$$f_C^{app} = F_C^{app}/F_R = (e^3/4\pi\varepsilon_0)\phi^{-2} F_C^{app} \cong (1.439965 \text{ eV}^2 \text{ V}^{-1} \text{ nm}) \cdot \phi^{-2} F_C^{app}. \quad (7)$$

Here, universal constants are stated to 7 significant figures, but should be rounded as appropriate. For a $\phi$=4.50 eV emitter, $F_R$ is about 14.1 V/nm.)

Alternatively, the procedure described in Ref. 3 can be used to make a direct extraction of apparent scaled-field values from a published data-analysis plot.

The question now is what the highest value of characteristic local field $F_C$ at which the emitter can be run stably, without tending to self-destruct as result of excessive heating or some other effect.

It is clear that the reference field should be an upper bound on this "breakdown field". If the top of the barrier is at the Fermi level, then many emitter electrons will rapidly escape and this will lead to emitter self-destruction, either due to Nottingham and/or Joule heating or (possibly) due to Coulomb explosion between incompletely screened ions in the apex region of the emitter.

Thus, if analysis, as above, of the published data appears to indicate that an emitter is operating well above the reference field (or, equivalently, well above $f_C^{app} = 1$), then the authors are reporting a "magic emitter". The reality, of course, is that their FE system is not electronically ideal, that their data analysis is faulty, and their published apparent field enhancement factor values are spurious.

In fact, electrical breakdown can occur at much lower fields. For tungsten emitters in a traditional field electron microscope configuration, Dyke and Trolan[17,18] assess that the safe continuous-current operating field is equivalent to $f_C$=0.34, which (for a $\phi$= 4.50 eV emitter) is equivalent to around 4.8 V/nm. Other experimental arrangements might have slightly higher breakdown fields.

If we use the same (slightly arbitrary) criteria as in the Orthodoxy Test, then (for a $\phi$= ~4.50 eV emitter):

- the published apparent MFEF value is *unreliable* if any part of the derived operating range of apparent scaled-field values has $f_C^{app} > 0.45$;

– the published apparent MFEF value is *almost certainly spurious* if any part of the derived operating range of apparent scaled-field values has $f_C^{app} > 0.75$.

For a 4.50 eV emitter, the corresponding apparent local field values are $F_C^{app} \sim 6.3$ V/nm and $F_C^{app} \sim 10.5$ nm V/nm, respectively.

## V. THE NEED FOR METHODS OF DIAGNOSIS

A current best guess, based on a small survey[3], is that up to 40% of the many hundreds of published values of apparent field enhancement factors may be spurious, in the sense that the reported experimental data would fail a validity check.

Significant questions are how to interpret this "failed" experimental data, or (in some cases) how to "adjust" a spurious derived LVCL or MFEF value, in order to derive at least a rough estimate of what the "true" value would be if the FE system were electronically ideal.

There exists a method of the latter kind, called *phenomenological adjustment.* This has been described elsewhere[19] and will not be discussed in detail here.

To interpret experimental data that has failed a validity check, it would usually be necessary to know the cause of failure. But, as noted earlier, there can in principle be many different causes of failure, and more than one cause could be operating.

In general, making an accurate diagnosis of the cause(s) of validity-check failure is an unsolved problem in the theory of how to interpret measured current-voltage data, and an unsolved problem more generally in FE systems engineering. In general, the problem is complicated and difficult, and deserves much more research attention than it has so far received. Some initial suggestions follow about what might be possible ways forward.

(*1) Measure the total energy deficit (TED) of an electron emerging from the emitter apex, either by retarding potential analysis (e.g., Refs 20,21), or by a direct contact method (e.g., Ref. 22). Measurement of a TED value of a few eV (or slightly more) may be an indicator that voltage-loss effects along the emitter are reducing the MFEF–value[13].

(*2) As suggested by Bachmann et al.[23], model the system as a resistance $Z$ in series with a field emitter, and establish the functional dependence of $Z$ on measured current and on any other relevant parameters.

(*3) Try to establish if physically different causes lead to qualitatively different shapes for [$I_m(V_m)$] data plots or related FN (etc.) plots.

(*4) In cases where the low-voltage part of a FN or related data-plot exhibits orthodox behaviour, but the high-voltage part "saturates", investigate if it might be possible to obtain useful information from study of how the "observed" slope correction factor varies with $V_m$, for high $V_m$.

## VI. OTHER SUGGESTIONS

Eventually, it may be useful to develop a short course on FE Systems Engineering and the interpretation of measured FE current-voltage characteristics. Some suggestions for content are made in the Appendix.

It also seems that it might be useful to modify the formal scope of the International Vacuum Nanelectronic Conference (and of related special issues of JVSTB). Thus the " Science and Technology" section of the IVNC "Call for Topics" could perhaps be modified in the following way (additions in italics):

"Science, *Engineering* and Technology

- *Emission* **physics** – Electron emission theory, including ab intitio and classic tunneling approaches.

- ***Emission* modeling** – modeling and simulation of electron emission physics from surfaces and devices, including microtips, nanogaps, photoemission, etc.

- ***Field Emission Systems Engineering** – Theory and electrical-engineering analysis of complete field electron emission systems, including methods for interpreting measured current-voltage characteristics.*

- **Fundamental** – Fundamental studies …

As already indicated, it seems to the author that if we wish to put the subject area of field electron emission onto a better and more respectable scientific basis, then a key requirement is for the subject area to carry out much more research into methodologies for interpreting measured current-voltage characteristics, particularly those taken from FE systems that are not electronically ideal. Hence this suggestion that we should explicitly make this a topic of stated interest to IVNC.

## VII.  SUMMARY

It has been argued that the interpretation of FE measured current-volt characteristics should be regarded as part of a specialized branch of Electrical/Electronic Engineering, called here "Field Electron Emission Systems Engineering". Its two major components would be: (a) emission physics; (b) "integrating theory" (including electrical circuit theory) that builds relationships between voltages, fields and currents. Field emitter electrostatics can be seen as auxiliary topic that contributes to both components, though more to the integrating theory than to the emission physics.

FE Systems Engineering needs to include validity checks that can determine whether a FE system is "electronically ideal" (i.e., current-voltage characteristics are determined by emission physics and zero-current system electrostatics alone, and hence can be validly interpreted by modern versions of traditional methodology, such as Fowler-Nordheim plots). Twelve different effects/situations that can cause an FE system to be electronically *non*-ideal have been identified in this article.

There is a need to establish methodologies that can diagnose the cause or causes of non-ideality. Some suggestions about possible methodologies have been made, but there is a great need for systematic research into this issue, and into the wider general issue of how to set about interpreting non-ideal current-voltage characteristics. It has been argued that the importance of these issues should be recognized by modifying the "calls for topics" for IVNC and for the related special issue of JVSTB.

It has also been proposed that it might be useful to develop a short course on FE Systems Engineering. Suggestions have been made about possible content.

It can be argued that, in order to be regarded as scientifically respectable, a research topic needs to have reliable procedures for making comparisons between theory and experiment. Currently, this aspect of field electron emission is undesirably weak, both for reasons discussed above and for more basic reasons[24]. Introducing the idea of FE Systems Engineering could be a useful part of the process of putting field electron emission onto a better and more respectable scientific basis.

## APPENDIX: POSSIBLE SHORT-COURSE CONTENT

Basic knowledge needs for FE Systems Engineering include the following topics. Ideally, most of these could be included in a short course.

– Relevant basic aspects of statistical thermodynamics and of the theory of electricity.

– Relevant aspects of the theory of charged surfaces, and of work-functions & related concepts.

– Planar emission physics, at least at the level of so-called "21$^{st}$ Century smooth-surface planar FE theory" (see arXiv 2107.08801).

– As relevant, theory relating to the effect of temperature on FE.

– Field emitter electrostatics, in the zero-current approximation.

– Introduction to finite-current FE-system electrostatics, and related practical implications.

- Introduction to the physics of field emitted vacuum space-charge (FEVSC).

- Theory and phenomenology of field-induced stress (Maxwell stress).

- Introduction to the dynamics of adsorption, migration and desorption of surface adsorbates, and to the role of adsorbates in emission physics.

- Electronic/electrical circuit modelling for FE systems, including (where appropriate and available) relevant forms of equivalent circuit.

- Standard "planar emission theory" methodologies for interpreting measured current-voltage data taken from electronically ideal systems.

- Methodology for applying "earthed-sphere-model" data-analysis techniques, using the Kryritsakis-Xanthakis FE equation[6] and/or a related procedure[5].

- As relevant, introductions to the special problems of FE from: (a) semiconductors; (b) carbon nanotubes; (c) graphene edges and blade-type emitters; and (d) rough and/or inhomogeneous surfaces.

- Introduction to relevant atomic level and bond-level emission physics, as these things may affect FE theory.

- Introduction to the "deep problems" of FE, that relate to the issue to how to apply quantum mechanics to FE in a manner that is "exactly correct".

# AUTHOR DECLARATIONS

## Conflict of interest

The author has no conflicts to disclose.

# DATA AVAILABILITY

Data sharing is not applicable to this article as no new data were created or analyzed in this study.